\documentclass[aps,physrev,twocolumn,10pt,superscriptaddress]{revtex4-2}

\usepackage[utf8]{inputenc}
\usepackage[english]{babel}
\usepackage[T1]{fontenc}
\usepackage{color}
\usepackage{xcolor}
\usepackage{microtype}
\usepackage{enumitem}
\usepackage[normalem]{ulem}
\usepackage[pdftex]{graphicx}
\usepackage{hyperref}
\usepackage{amsmath}
\usepackage{amssymb}
\usepackage{mathtools}
\usepackage{mathrsfs}
\usepackage{multirow}
\usepackage{bbm}
\usepackage{bbold}
\usepackage{braket}
\usepackage[version=4]{mhchem}
\usepackage[only,llbracket,rrbracket]{stmaryrd}
\usepackage{siunitx}
\usepackage[caption=false]{subfig}
\usepackage{float}
\usepackage{makecell}
\usepackage[percent]{overpic}
\usepackage[capitalize]{cleveref}

\hypersetup{%
    hypertexnames=false,
    colorlinks=true,
    allcolors=blue,
    unicode=true
}

\newcommand{\phantomsubfloat}[1]{%
    {%
        \captionsetup[subfloat]{farskip=0pt,captionskip=0pt}
        \captionsetup[subfigure]{labelformat=empty}
        \subfloat{#1}
    }
}

\crefname{section}{Sec.}{Sec.}
\crefname{appendix}{Appendix}{Appendixes}

\begin{document}

\title{Optimized measurement-free and fault-tolerant quantum error correction \texorpdfstring{\\}{} for neutral atoms}

\author{Stefano Veroni}
\email{stefanoveroni00@gmail.com}
\affiliation{PlanQC GmbH, M\"unchener Str. 34, 85748 Garching, Germany}

\author{Markus M\"uller}
\affiliation{Institute for Quantum Information, RWTH Aachen University, D-52056 Aachen, Germany}
\affiliation{\mbox{Peter Gr\"unberg Institute, Theoretical Nanoelectronics, Forschungszentrum Jülich, D-52425 Jülich, Germany}}

\author{Giacomo Giudice}
\affiliation{PlanQC GmbH, M\"unchener Str. 34, 85748 Garching, Germany}

\date{\today}

\begin{abstract}
    A major challenge in performing quantum error correction (QEC) is implementing reliable measurements and conditional feed-forward operations.
    In quantum computing platforms supporting unconditional qubit resets, or a constant supply of fresh qubits, alternative schemes which do not require measurements are possible.
    In such schemes, the error correction is realized via crafted coherent quantum feedback.
    We propose implementations of small measurement-free QEC schemes, which are fault-tolerant to circuit-level noise.
    These implementations are guided by several heuristics to achieve fault-tolerance: redundant syndrome information is extracted, and additional single-shot flag qubits are used.
    By carefully designing the circuit, the additional overhead of these measurement-free schemes is moderate compared to their conventional measurement-and-feed-forward counterparts.
    We highlight how this alternative approach paves the way towards implementing resource-efficient measurement-free QEC on neutral-atom arrays.
\end{abstract}

\maketitle

\section{Introduction}
\label{sec:introduction}
Large-scale quantum computers will require \emph{quantum error correction} (QEC) to perform quantum information processing~\cite{terhal2015,campbell2017}.
By carefully designing the physical system one can counteract the propagation of faults, arising from fundamentally noisy hardware.
The logical information is encoded in a small subspace of the physical Hilbert space of the whole system, in a typically non-local way: this \emph{redundant} encoding is then robust against local errors.
The redundancy however introduces a significant overhead in the resources required to implement QEC.
Therefore, the choice of the specific error-correction scheme should be tailored to a given quantum computer, in order to minimize this overhead and take advantage of the features of a given physical platform.
Despite the technical challenges, we are currently witnessing exciting progress in the field of QEC, as highlighted by recent experimental advances~\cite{egan2021, google2021, google2023, ryan-anderson2021, krinner2022, abobeih2022, zhao2022, wang2023, gupta2024, hilder2022, postler2022, postler2023, bluvstein2023, huang2023, pogorelov2024, dasilva2024}.

\begin{figure}[!ht]
    \begin{overpic}[width=\linewidth]{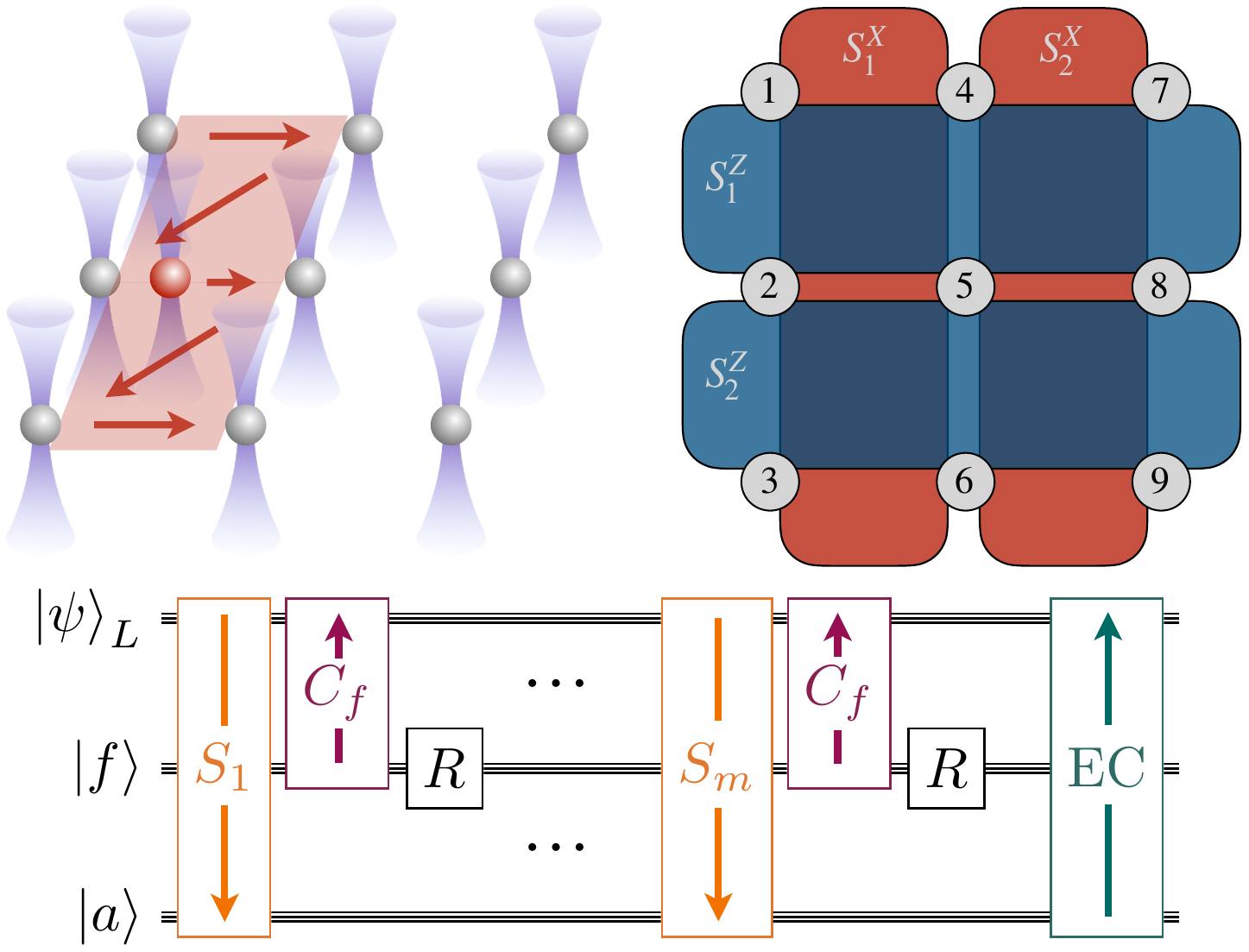}
        \put(6, 65){(a)}
        \put(52, 65){(b)}
        \put(6, 25){(c)}
    \end{overpic}
    \vspace*{-2em}
    \phantomsubfloat{\label{fig:intro-tweezers}}
    \phantomsubfloat{\label{fig:codes-bacon-shor}}
    \phantomsubfloat{\label{fig:intro-qec-circ}}
    \caption{
        (a) Optical tweezer arrays allow for the manipulation of individual atoms in space.
        By placing atoms nearby, one can exploit Rydberg interactions to perform entangling gates.
        Such gates can then be used to extract stabilizer information from the system.
        For example, the ancillary qubit highlighted in red can be used to extract the stabilizer $S_1^X = X_1 X_2 X_3 X_4 X_5 X_6$ of the Bacon-Shor code.
        (b) The Bacon-Shor code-space is generated by four six-body stabilizers: two $X$-stabilizers in the vertical direction and two $Z$-stabilizers in the horizontal direction.
        (c) General circuit layout for fault-tolerant measurement-free QEC. 
        It uses three registers of qubits: data $\ket{\psi}_L$, flags $\ket{f}$, ancillae $\ket{a}$.
        Fault-tolerance is achieved via two main ingredients: syndrome redundancy and single-shot flags.
        The circuit coherently maps syndrome information of a redundant number $m$ of stabilizers on ancilla qubits, with $m$ larger than the size of the generating set of stabilizers (orange).
        Each extraction is aided by flags, which detect errors occurring during the coherent mapping of the stabilizers. 
        The flag pattern is then used to correct these errors on data qubits (purple).
        After the extraction of each stabilizer, the qubits in the intermediary register are reset to $\ket{0}$ or, alternatively, replaced by fresh qubits in $\ket{0}$ ($R$ gate).
        Finally, the correction is applied coherently based on the syndrome information contained in the quantum state of the ancillae (green). 
        Arrows represent the flow of information.
    }
    \vspace*{-3em}
    \label{fig:intro}
\end{figure}

Neutral atom arrays are a promising candidate for large-scale quantum information processors~\cite{henriet2020}.
Atoms can be manipulated in large numbers using optical tweezers, cf.~\cref{fig:intro-tweezers}, enabling reconfigurable geometries as well as the physical transport of individual atoms while preserving quantum information~\cite{bluvstein2022}.
The quantum information is stored in different atomic levels, exhibiting long coherence times of the order of seconds.
Single-qubit gates can be realized with controlled sequences of microwave or laser pulses, which can be performed in times of the order of $\SI{0.5}{\micro\second}$~\cite{levine2022,ma2022,unnikrishnan2024,pucher2024}.
Multi-qubit entangling gates are realized via Rydberg interactions between neighboring qubits.
By engineering pulses that drive the atoms to their Rydberg states and back~\cite{jaksch2000,levine2019}, high-fidelity controlled-$Z$ gates ($CZ$) have been realized with rubidium~\cite{evered2023}, strontium~\cite{cao2024, finkelstein2024}, ytterbium~\cite{peper2024}, and cesium~\cite{radnaev2024}.
Furthermore, these gates can be slightly faster than single-qubit gates.
One can generalize this mechanism to natively realize multi-controlled gates as well, such as $C^m Z$, which were demonstrated experimentally~\cite{levine2019,evered2023}.
This is particularly interesting from an algorithmic point of view, as this enables fast Toffoli ($CCX$) gates, which can be constructed from $CCZ$ and local Hadamard ($H$) gates. 

One limitation of neutral-atom platforms is state measurement, which is typically performed by inducing fluorescence and detecting the light emitted from the atoms.
Feed-forward operations based on real-time measurements remain challenging, despite recent demonstrations~\cite{ma2023,singh2023,huie2023,graham2023,lis2023,norcia2023,bluvstein2023}.
Measurements require times of the order of $\SI{500}{\micro\second}$, without accounting for camera readout times and possible shuttling time of atoms to a dedicated readout zone~\cite{shea2020, bluvstein2022, bluvstein2023}.
Therefore, there are around three orders of magnitude in time difference between gates and measurements in neutral-atom platforms.
Coupling the atoms to a cavity could speed up the times required for read out~\cite{bochmann2010,deist2022}, at the cost of a loss of parallelism for single-mode cavities. 

In standard QEC protocols, some partial information of the system is measured, the \emph{error syndrome}.
Based on this, a correction is decided upon by a classical algorithm (decoder) and applied to the system, either by physical feedback or in software~\cite{terhal2015}.
Relatively slow measurements inhibit the fast execution of such protocols.
As an alternative, \emph{measurement-free} (MF) QEC protocols have been proposed~\cite{paz-silva2010,li2012,crow2016,ercan2018}, where the classical processing is replaced with unitary dynamics.
The overall dynamics are still dissipative, since we need to remove the additional entropy introduced by faults.
In MF QEC, this is provided by reset operations, or a sufficiently large reservoir of continuously-provided ancilla qubits~\cite{singh2022, norcia2024, gyger2024}.

Recent attempts have focused on rendering such MF schemes fault-tolerant (FT): such FT constructions guarantee that the protocol is capable of lowering failure rates provided the physical error rates are below a break-even point, by introducing redundancies in the syndrome under a specific error model~\cite{perlin2023}.
The first MF and fully FT explicit quantum circuit implementation was proposed in Ref.~\cite{heussen2023a}, making use of concepts from Steane-type error correction \cite{steane1997} (not to be confused with the Steane code, discussed later).
There, the key is to first fault-tolerantly copy the stabilizer information of the to-be-corrected logical data qubit to an auxiliary logical qubit register to avoid the uncontrolled propagation of errors to the logical data qubit during the extraction of the error syndrome. 
This additional register needs to be initialized fault-tolerantly into a logical state to perform the coherent error copying-operation, which requires further overhead.

In this paper, we develop alternative strategies to construct MF, yet fully FT QEC circuits, which are ideally suited for application to low-distance QEC codes such as, e.g., the nine-qubit Bacon-Shor code illustrated in \cref{fig:codes-bacon-shor}.
As sketched in \cref{fig:intro-qec-circ}, these protocols involve (i) introducing redundancies in the syndrome information, and (ii) using additional qubits to first flag the syndrome extraction and to then apply corrections on the data qubits in a single-shot fashion.
These procedures reduce the qubit overhead, and, simultaneously, avoid the need for fault-tolerant preparation of auxiliary logical states.
In \cref{sec:guidelines}, we outline our strategy, and formulate some general heuristics with which we design concrete error-correcting circuits.
In \cref{sec:codes} we demonstrate the validity of our scheme by applying them to specific distance-three QEC codes, namely, the Bacon-Shor code~\cite{bacon2006, aliferis2007}, Shor's code~\cite{shor1995}, the rotated surface code~\cite{bravyi1998, dennis2002, fowler2012} and Steane's code~\cite{steane1996} as the smallest two-dimensional (2D) topological color code~\cite{bombin2006}. 
The performance of our MF QEC protocols is then benchmarked and compared to comparable feed-forward implementations in \cref{sec:performance}, for a general depolarizing noise model as well as a noise model specific to neutral-atom arrays.
Finally, we summarize our results and discuss the possibilities of scaling up such approaches in \cref{sec:conclusion}.

\section{Guidelines for fault-tolerant measurement-free QEC}
\label{sec:guidelines}
A QEC code is described by the set of integers $\llbracket n, k, d \rrbracket$, where $n$ is the number of physical qubits, $k$ is the number of logical qubits and $d$ is the code distance, meaning the code can correct at least $t = \left \lfloor{(d-1)/2}\right \rfloor$ errors.
A protocol is said to be FT if it is capable of lower logical failure rates than those of the physical qubit operations, provided that the latter are below a break-even point~\cite{aharonov1998, preskill1998}.
This is the case when, if the input is a codeword with error of weight $r$ and the protocol has $s$ faults, with $r+s\leq t$, then the output is the original codeword up to a correctable error.
We consider general circuit-level error models, which correspond to some (possibly gate-dependent) quantum channel following each operation.
We therefore aim at designing specific circuits implementing QEC protocols, which are FT against such errors.

Recently, fault tolerance was achieved by adopting elements of Steane-type error correction, by introducing an auxiliary register of qubits to be used as intermediary between data and ancillae \cite{heussen2023a}.
There, errors on data qubits are copied coherently to an auxiliary register by means of a transversal $\mathrm{CNOT}$.
This auxiliary register is then manipulated during the syndrome extraction, such that errors occurring in this register do not propagate to the data qubits.
In other words, the lack of direct communication between data qubits and ancillae enables fault tolerance.

In this work, we suggest an alternative scheme, which allows one to use fewer qubits and operations, as well as gates with support on fewer qubits (at most 3-qubit Toffoli-type gates), to achieve similar performances.
As summarized in~\cref{fig:intro-qec-circ}, our strategy is built upon the following elements:
\begin{enumerate}
    \item \emph{Syndrome redundancy}.
    An informationally over-complete set of stabilizer bits is extracted \cite{crow2016}.
    This prevents that errors during the syndrome extraction lead to erroneous corrections, corrupting the logical state.
    It also allows for the application of the aforementioned heuristic to be applied in the error correction.
    The order in which the stabilizers are read out needs to be chosen carefully to satisfy certain constraints, as explained in \cref{appendix:redundant permut}.
    
    \item \emph{Single-shot flags}.
    Additional qubits are used to flag the syndrome extraction~\cite{chao2018}.
    Flag qubits prevent hook errors, i.e. mid-extraction bit-flips on the ancillae, from inducing uncorrectable errors on the data.
    Typically, if a flag qubit is triggered, i.e.~measured in the $\ket{1}$ state, the syndrome extraction circuit is repeated.
    Our use of flag qubits is single-shot~\cite{prabhu2023}, as they lead to corrections on the data qubits, immediately after each stabilizer read out, without the need for repeating the extraction circuit.
    Notably, we show this can be done without measurements.
\end{enumerate}
In the next section, we provide explicit examples how these guidelines are implemented.
As will be shown, they allow for single-round FT QEC, where only one round of stabilizer measurements is required.

\begin{figure}[t]
    \begin{overpic}[width=\linewidth]{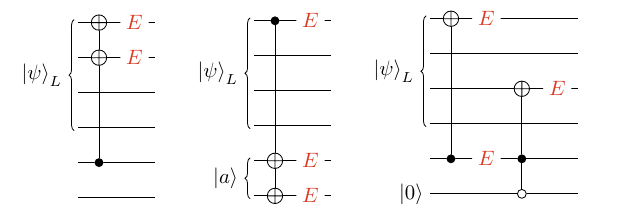}
        \put(2, 32){(a)}
        \put(30, 32){(b)}
        \put(60, 32){(c)}
    \end{overpic}
    \vspace{-1.5em}
    \caption{
        Configurations or circuit elements \emph{to be avoided} to maintain fault-tolerance in a $d=3$ QEC protocol.
        (a) Gates involving more than one data qubit, as they can introduce errors on several data qubits.
        (b) Gates involving a data qubit and more than one ancilla. In fact these can immediately cause an error on the data qubit.
        Additionally, multiple errors on the ancillae introduce additional errors on the data later in the circuit.
        In contrast, single errors on the ancillae can be tolerated by imposing a minimal syndrome redundancy, discussed in the main text.
        (c) Multi-controlled gates acting on data qubits with common one-controls can cause errors on multiple data qubits. 
        This can occur if all one-controls of the later gate are shared with the earlier gate.
        The different treatment of zero-controls and one-controls derives from the fact that the qubits they are controlled on are expected to be in zero in absence of errors, as for an ancilla qubit.
        Then, as depicted, a single error activates the latter multi-controlled gate. 
    }
    \label{fig:guidelines}
\end{figure}

While the general strategy outlined above enables fault-tolerance, it does not automatically provide FT circuit implementations.
For this goal, we suggest that a FT $d = 3$ QEC protocol should avoid the following configurations (illustrated in \cref{fig:guidelines}):
\begin{enumerate}[label=(\alph*)]
    \item multi-qubit gates involving more than a single data qubit;
    \item multi-qubit gates in the syndrome extraction acting on a data qubit and more than one ancilla;
    \item certain pairs of multi-qubit gates acting on data and sharing some qubits. 
    Specifically, those outside of the syndrome extraction, where all the one-control qubits~\footnote{%
        In a controlled unitary, we define a one-control as the qubit that leads to the application of the unitary if the control is in $\ket{1}$.
        These are illustrated in the diagrams with a filled circle.
    }
    of a later gate are also involved in an earlier one.
\end{enumerate}
As illustrated in \cref{fig:guidelines}, these guidelines prevent the spread of errors on the input logical qubit, under the assumption of correlated errors occurring at first-order in probability on the qubits of a single gate.
These guidelines are neither sufficient on their own, nor strictly necessary.
In fact, they could be relaxed on a case-by-case basis, depending on the presence of correlation or bias in the noise, additional structure of the QEC code, or the specific syndrome redundancy.
Nevertheless, they can aid in the realization of generally FT QEC implementations.

\begin{figure*}[t]
    \includegraphics[width=\textwidth]{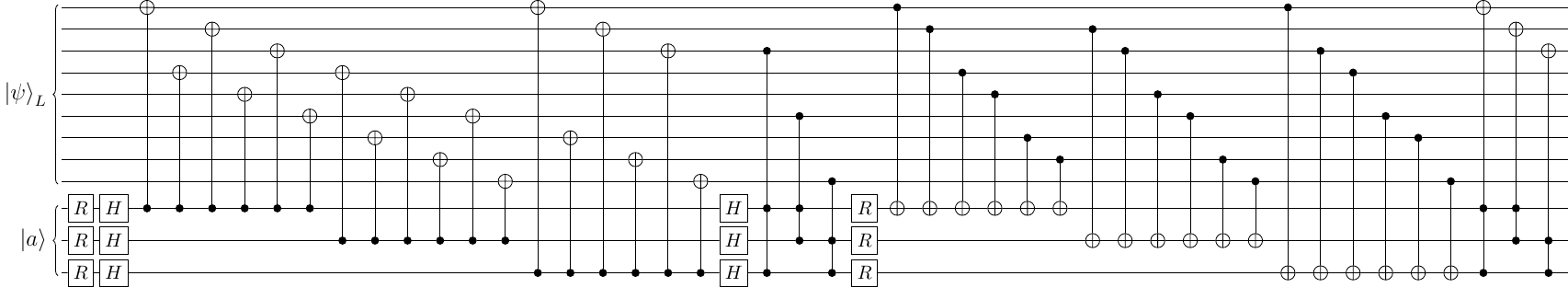}
    \vspace{-1.5em}
    \caption{
        Measurement-free fault-tolerant QEC implementation for the Bacon-Shor code. 
        It coherently reads out three $X$-type and $Z$-type stabilizers rather than only two. 
        This redundant syndrome information allows for fault-tolerant measurement-free error correction via coherent quantum feedback, here by Toffoli-type three-qubit gate operations.
        The left half of the circuit detects and corrects $Z$-errors; the right half, after the reset $R$ of the ancillae to the $\ket{0}$ state, detects and corrects $X$-errors.
        In this scheme, no flag qubits are required, as hook errors correspond to single-qubit errors up to a change of gauge.
    }
    \label{fig:bacon-shor}
\end{figure*}

\section{Fault-Tolerant Measurement-Free Protocols}
\label{sec:codes}

In this section, we outline the FT circuit implementation for the Bacon-Shor code and Shor's code. 
These will illustrate our guidelines presented in \cref{sec:guidelines}.

These guidelines can be easily extended to any Calderbank-Shor-Steane (CSS) code~\cite{calderbank1996, steane1996-CSS}.
In \cref{appendix:surf-steane}, we provide additional circuit implementations for the surface code and Steane's code.
In a CSS code, stabilizers are either $X$-type or $Z$-type Pauli strings.
Logical Pauli operators are transversal, and so is the logical $\mathrm{CNOT}$.
These codes allow one to treat bit and phase flips with separate sub-circuits, referred to as $X$- and $Z$-blocks, and which are executed one after the other.

The general procedure for a MF implementation requires the following steps: (i) determining the most adequate form of redundancy in the syndrome extraction, (ii) possibly placing flag operations  to detect and correct certain errors occurring in stabilizer extractions, and (iii) designing a correction circuit benefiting from the redundancy. 
We expect that this strategy can be further extended to any stabilizer code, albeit with a higher overhead.

\subsection{Bacon-Shor Code}
\label{sec:bacon-shor}

The smallest Bacon-Shor code~\cite{bacon2006, aliferis2007} encodes one logical qubit in nine physical qubits. 
Its code-space is the joint $+1$ eigenspace of the following set of mutually commuting stabilizers generators 
\begin{equation}
    \begin{aligned}
        S_1^X &= X_1 X_2 X_3 X_4 X_5 X_6, \quad S_1^Z &= Z_1 Z_2 Z_4 Z_5 Z_7 Z_8, \\
        S_2^X &= X_4 X_5 X_6 X_7 X_8 X_9, \quad
        S_2^Z &= Z_2 Z_3 Z_5 Z_6 Z_8 Z_9, 
    \end{aligned}
    \label{eq:bacon-shor-stabilisers}
\end{equation}
as shown in~\cref{fig:codes-bacon-shor}.
Its logical operators can be chosen to be $X_L = X_1 X_2 X_3$ and $Z_L = Z_1 Z_4 Z_7$.
The code supports a transversal logical Hadamard gate up to a qubit permutation \cite{aliferis2007}, and can be fault-tolerantly encoded with no additional overhead~\cite{egan2021}.

In the Bacon-Shor code, which is a subsystem code~\cite{terhal2015}, multiple states can correspond to the same codeword.
These states are equivalent up to gauge operators, which commute with both the logical operators and the stabilizers, and as such do not affect the logical information. 
The group of gauge operators is generated by the pairs $X_i X_j$ ($Z_i Z_j$) of elements $i,j$ in the same row (column) as of~\cref{fig:codes-bacon-shor}.
As the gauge is allowed to change, multiple pairs of operations $X_i X_j$ and $Z_i Z_j$ do not constitute an error. 

The addition of syndrome redundancy is straightforward. 
Aside from the stabilizers in \cref{eq:bacon-shor-stabilisers}, we also extract the eigenvalues of
\begin{equation}
    \begin{aligned}
        S_3^X &= S_1^X S_2^X = X_1 X_2 X_3 X_7 X_8 X_9, \\
        S_3^Z &= S_1^Z S_2^Z = Z_1 Z_3 Z_4 Z_6 Z_7 Z_9.
    \end{aligned}
\end{equation}
These preserve the separation between $X$- and $Z$-blocks.
Due to the resilience of the code to multiple higher-weight errors, the extraction of stabilizers does not require flag qubits. 
In fact, by suitably ordering extraction operations, any hook error results in at most one faulty data qubit, up to a gauge operator.

Finally, the correction blocks use Toffoli gates to apply corrections on data qubits, as shown in the circuit in ~\cref{fig:bacon-shor}.
Each Toffoli is activated by the pair of faulty syndromes associated with that particular error.
The positioning of the gates is done in accordance with the heuristics outlined in the previous section.
In particular, the redundancy in the syndromes allows the gates to avoid a full overlap of controls.
In turn, this prevents single faults on the ancillae leading to erroneous correction operations, and hence uncorrectable errors.

\subsection{Shor's Code}
\label{sec:shor}

\begin{figure*}[t]
    \begin{overpic}[width=\linewidth]{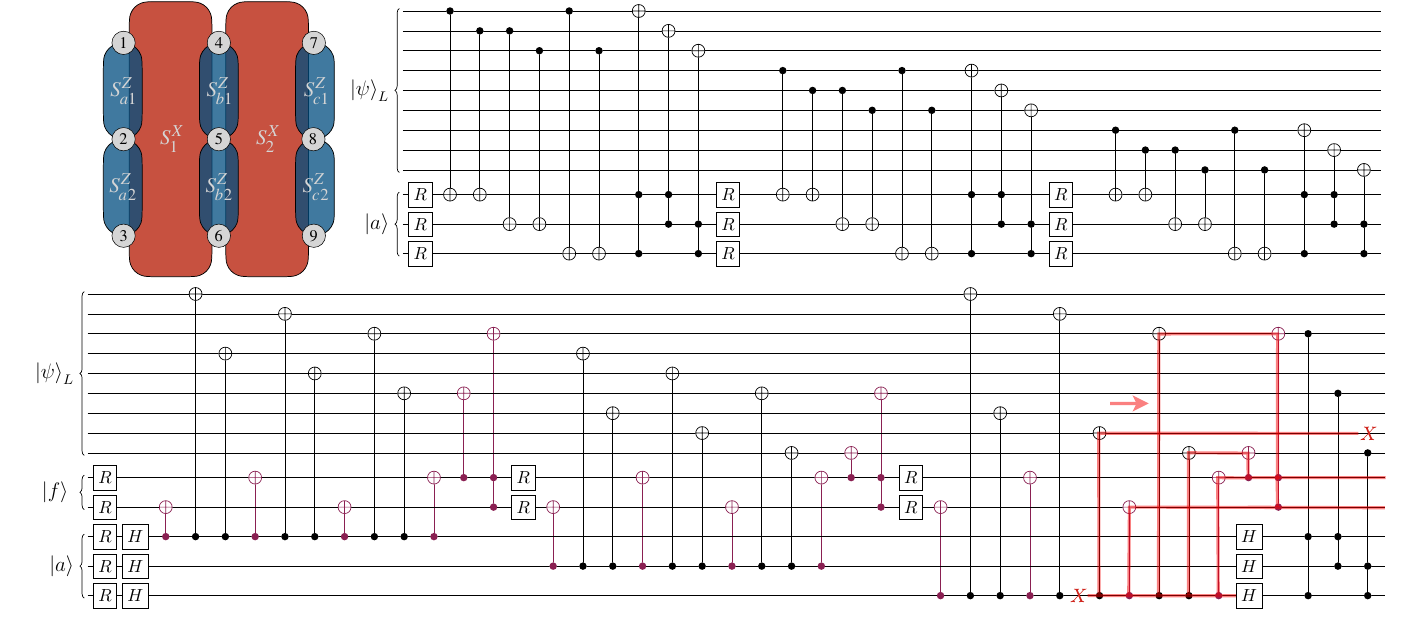}
        \put(2, 43){(a)}
        \put(25, 43){(b)}
        \put(2, 23){(c)}
    \end{overpic}
    \phantomsubfloat{\label{fig:shor-stabilizers}}
    \phantomsubfloat{\label{fig:shor-Xblock}}
    \phantomsubfloat{\label{fig:shor-Zblock}}
    \vspace{-1.5em}
    \caption{
        Measurement-free fault-tolerant QEC implementation for the nine-qubit Shor code.
        (a) Stabilizers of the Shor's code. 
        The $S^Z$ stabilizers can be grouped into pairs acting on different triplets of qubits (a pair for each column).
        (b) Circuit for the correction of bit flips. 
        Note how the syndrome extraction and correction occur separately for each triplet of qubits having separate $S^Z$ stabilizers.
        (c) Circuit for the correction of phase flips.
        Operations acting on flags are depicted in purple, and their specific location is chosen to detect and correct uncorrectable hook errors.
        The two flag operations at the end of each stabilizer measurement (a CNOT and a Toffoli) apply these corrections on the data qubits.
        An example of how a hook error propagates and is caught by the flag qubits is highlighted in red.
    }
    \label{fig:shor}
\end{figure*}

Shor's code \cite{shor1995} is a concatenation of the bit-flip and phase-flip codes. 
It encodes one logical qubit in nine physical qubits. 
Its code space is the joint $+1$ eigenspace of the following stabilizer generators 
\begin{equation}
    \begin{aligned}
    S_1^X &= X_1 X_2 X_3 X_4 X_5 X_6, \\
    S_2^X &= X_4 X_5 X_6 X_7 X_8 X_9, \\
    S_{a1}^Z &= Z_1 Z_2, \\
    S_{a2}^Z &= Z_2 Z_3, 
    \end{aligned}
    \qquad \;
    \begin{aligned}
    S_{b1}^Z &= Z_4 Z_5, \\
    S_{b2}^Z &= Z_5 Z_6, \\
    S_{c1}^Z &= Z_7 Z_8, \\
    S_{c2}^Z &= Z_8 Z_9,
    \label{eq:shor stabilizers}
    \end{aligned}
\end{equation}
as shown in~\cref{fig:shor-stabilizers}.
Its logical operators can be chosen to be $X_L = X_1 X_2 X_3$ and $Z_L = Z_1 Z_4 Z_7$.
Encoding the $\ket{\pm}_L$ codeword is trivially FT.
As evident in ~\cref{fig:shor-stabilizers}, the $S^Z$ stabilizers can be grouped into pairs acting on separate triplets of qubits (labeled in~\cref{eq:shor stabilizers} by letters $a,b,c$).
Therefore, they can be read out independently, allowing one to correct single bit-flip errors acting on different triplets of qubits, e.g. $X_2 X_5$ or $X_1 X_4 X_9$.

The most adequate syndrome redundancy is the one preserving the structure of the code.
In order to maintain the separation between $X$- and $Z$-blocks, and between the triplets with different $S^Z$ stabilizers, we also extract the following stabilizers
\begin{equation}
    \begin{aligned}
    S_3^X &= X_1 X_2 X_3 X_7 X_8 X_9, \\
    S_{a3}^Z &= Z_1 Z_3,  
    \end{aligned}
    \qquad \;
    \begin{aligned}
    S_{b3}^Z &= Z_4 Z_6, \\
    S_{c3}^Z &= Z_7 Z_9,
    \label{eq:shor add stabilizers}
    \end{aligned}
\end{equation}

Read-out of weight-two stabilizers does not require flags. 
Instead, the read-out of weight-six stabilizers is made FT by the addition of two flag qubits.
As illustrated in \cref{fig:shor-Zblock}, these flags detect bit-flips on the ancilla occurring during the extraction. 
The placement of the flag operations is chosen to detect hook errors~\cite{prabhu2023}.
Then, gates controlled on the flags and acting on the data prevent said bit-flips from becoming unrecoverable errors, in a FT way.
In contrast to conventional flagged circuits \cite{chao2018}, this design does not require measurements, as the same extraction circuit is run independently of the values of the flags. 
This deterministic scheme resembles that in Ref.~\cite{prabhu2023}, but is simplified and implemented without measurements and feed-forward operations.
It requires fewer gates at the price of allowing multiple, yet correctable, bit-flips to propagate. 
This is not problematic as they can be corrected. 

Finally, the correction blocks closely resemble those of the Bacon-Shor code.
The full circuit of our implementation is shown in \cref{fig:shor}. 

\section{Performance}
\label{sec:performance}

In this section we benchmark our protocols against hardware-agnostic depolarizing noise (\cref{sec:depolarising-noise}), and a noise model tailored to neutral atom arrays (\cref{sec:biased-noise}). 
Details of the models are given in the relevant subsections.
We simulate a single round of error correction, as we are mostly interested in assessing the fault-tolerance of our protocols and comparing their relative performance, similarly to Ref.~\cite{heussen2023a}.
For estimating the performance of an actual device, it may be preferable to average the performance over multiple rounds of error correction, but we do not expect a qualitative difference.

We simulate our circuits, which heavily use non-Clifford gates, with state vector simulations using \texttt{cirq}~\cite{cirq}.
We note, however, that scalable methods to simulate MF QEC circuits have been developed, exploiting the fact that under a Pauli noise model the ancilla qubits are effectively classical bits~\cite{perlin2023}.
However, given that the system sizes are accessible to state vector simulations, we perform simulations with modest resources in the following fashion. 
First, we confirm the fault-tolerance of the protocols by simulation of all possible single faults.
To compute the logical failure probability, we perform a Monte Carlo simulation starting from an ideal codeword, sampling  Kraus operators at each error location according to the chosen noise model.
Because of fault-tolerance, we only simulate circuits with two or more errors.
The logical failure probability is then estimated as
\begin{equation}
    p_\mathrm{log} = p_\mathrm{log}^{2+} \cdot p_\mathrm{err}^{2+},
\end{equation}
where $p_\mathrm{log}^{2+}$ is the average failure probability associated to having two or more errors in the circuit, and $p_\mathrm{err}^{2+}$ is the probability two or more errors occurring.
This is given by
\begin{equation}
    p_\mathrm{err}^{2+} = 1 - p_\mathrm{err}^{0} - p_\mathrm{err}^{1},
\end{equation}
with
\begin{equation}
    p_\mathrm{err}^{0} = \prod_i (1 - p_i)^{N_i},  \quad
    p_\mathrm{err}^{1} = p_\mathrm{err}^{0} \sum_i \frac{N_i p_i}{1-p_i}, 
\end{equation}
where $p_i$ is the probability that a gate of type $i$ is faulty, and $N_i$ is the number of such gates.
The averaging is performed over $60\,000$ simulations, with inputs equally distributed among $\ket{0}_L$, $\ket{+}_L$ and $\ket{i}_L$~\footnote{%
    This corresponds to a Haar-uniform sampling over initial logical states.
}.

\subsection{Depolarizing Noise}
\label{sec:depolarising-noise}

\begin{figure*}[t]
    \vspace{1em}
    \begin{overpic}[width=\textwidth]{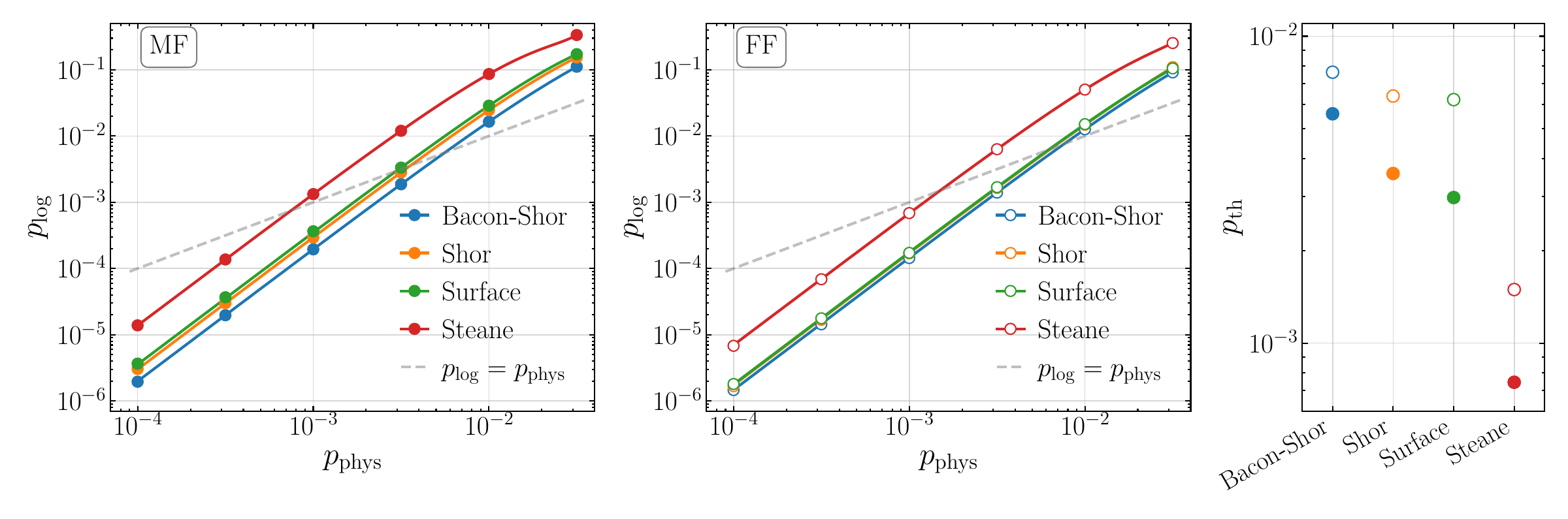}
        \put(0, 32){(a)}
        \put(40, 32){(b)}
        \put(77, 32){(c)}
    \end{overpic}
    \phantomsubfloat{\label{fig:FDN_MF}}
    \phantomsubfloat{\label{fig:FDN_FF}}
    \phantomsubfloat{\label{fig:FDN_pth}}
    \vspace{-2.5em}
    \caption{%
        Logical qubit error rate vs.\ physical one, under symmetric depolarizing noise of \cref{eq:FDN}, for the (a) measurement-free and (b) feed-forward implementations.
        Solid lines correspond to the numerical fits used to estimate the pseudo-thresholds. 
        The fit is performed with a polynomial of the form $p_\mathrm{log} = c_2 p_\mathrm{phys}^2 + c_3 p_\mathrm{phys}^3 + c_4 p_\mathrm{phys}^4$.
        Since single-qubit errors are correctable due to the FT character of the protocols, the zero-th and first order are vanishing, while terms beyond the fourth degree are negligible at the considered physical error rates.
        (c) Numerical estimates of the pseudo-thresholds.
    }
    \label{fig:plogs_FDN}
\end{figure*}

We consider the hardware-agnostic \emph{depolarizing channel}.
Given the set of Pauli strings of length $\ell$, i.e., $\mathcal{P}_\ell = \{I,X,Y,Z\}^{\otimes \ell}$, it is defined as 
\begin{equation}
    \mathcal{E}(\rho) = (1 - p) \rho + \frac{p}{|\mathcal{P_\ell}|-1} \sum_{\mathclap{\substack{P \in \mathcal{P}_\ell \\ P \neq I^{\otimes \ell}}}}{P \rho P^\dagger},
    \label{eq:FDN}
\end{equation}
where $p$ is the probability of an error occurring.
In other words, after each gate we apply with probability $p$ an operator sampled uniformly from all non-trivial Pauli strings.

The depolarizing channel is a fairly general error model, and is a standard benchmark of QEC protocols.
We study both a MF and feed-forward (FF) version of our QEC protocols. 
In the latter, flags and ancillae are measured, and the correction circuit is substituted by conventional correction operations, which are only applied if required according to a look-up table decoding. 
Initialization operations, measurements and correction operations have the same error rate $p_\mathrm{phys} = p$ as other operations, whereas idling errors are neglected.
The set of required gates $\{X, H, CX, CZ, CCX, CCZ\}$ is assumed to be natively supported.

The logical performance of each protocol is shown in \cref{fig:plogs_FDN}; their pseudo-thresholds and resource overhead are summarized in \cref{tab:summary-depolarizing}.
The pseudo-threshold is directly connected to the number of gates and the correction capabilities of each implementation.
It is not surprising that the implementations which are both shorter and resilient to higher weight errors---such as of the Bacon-Shor or Shor's code---lead to the best performance.
On the other hand, Steane's code cannot correct higher weight errors. 
Furthermore, the logic in its correction step is more involved and has the largest number of Toffoli gates.
As such, this protocol performs worse than other codes.
It is worth noting that both the correction capabilities and the resources overhead of a protocol can in part be inferred from the structure of the stabilizers.
In fact, codes with stabilizers generators having fewer common qubits, as in Shor's and the surface code, generally have lower logical error rates.

\begin{table}[t]
    \centering
    \def\arraystretch{1.4}
    \begin{tabular}{p{2.45cm}|c|c|c}
         Protocol          
         & \,\# qubits\, 
         & \makecell{\# gates \\ ($R$, $G_1$, $G_2$, $G_3$, $M$) }
         & \,$p_\mathrm{th} [\%]$
         \\
         \hline
         \hline
         Bacon-Shor (MF)   
         &  12  
         & \enspace6 \, \enspace6 \, 36 \, \enspace6 \, \enspace0   
         & 0.56
         \\
         Bacon-Shor (FF)   
         &  10  
         & \enspace6 \, \enspace6 \, 36 \, \enspace0 \, \enspace6   
         & 0.76
         \\
         \hline
         Shor (MF)         
         &  14  
         & 18 \, \enspace6 \, 51 \, 15 \, \enspace0  
         & 0.36
         \\
         Shor (FF)         
         &  12  
         & 18 \, \enspace6 \, 48 \, \enspace0 \, 18  
         & 0.64
         \\
         \hline
         Surface (MF)      
         &  17  
         & 20 \, 24 \, 40 \, 22 \, \enspace0 
         & 0.30
         \\
         Surface (FF)      
         &  10  
         & 12 \, 12 \, 40 \, \enspace0 \, 12 
         & 0.62
         \\
         \hline
         Steane (MF)       
         &  14  
         & 38 \, 26 \, 90 \, 32 \, \enspace0 
         & 0.07
         \\
         Steane (FF)       
         &  10  
         & 30 \, 20 \, 90 \, \enspace0 \, 30 
         & 0.15
    \end{tabular}
    \caption{%
    Comparison of measurement-free and feed-forward implementations, when natively supporting any required gate. 
    The number of qubits refers to the \emph{minimal} number required by the implementation, since each reset can be replaced by an additional qubit.
    The gate count is divided into initialization/reset, 1-, 2-, 3-qubit gates, and measurements. 
    For FF approaches, this does not count the Pauli corrections applied after the decoding of the measurements, as their number varies between runs.
    The pseudo-threshold $p_\mathrm{th}$ is computed under symmetric depolarizing noise in absence of idling errors.
    \label{tab:summary-depolarizing}
    }
\end{table}

Compared to the single-shot FF implementations, it is not surprising that the MF approach, which has considerably deeper circuits, returns a higher logical error rate at same physical error rate.
However, it is remarkable that the MF overhead consistently lowers the pseudo-threshold only by a factor of approximately 2.
This suggests that MF QEC represents a valuable alternative for architectures in which measurements are technically challenging or come with a large time overhead.
This will be explored in the next subsection.

We may also ask what is the performance of our single-shot FF implementations compared to conventional flag-based implementations in the literature.
For the Bacon-Shor code, Shor's code, and Steane's code, we consider the implementation with parallel flags outlined in Ref.~\cite{liou2023}.
Again, we stress that in this scheme, the flags are not single-shot, i.e., after they signal the presence of an error, a new circuit is run.
Overall, similar performance is achieved, since the pseudo-thresholds they obtained are $\SI{0.86}{\percent}$, $\SI{0.98}{\percent}$, and $\SI{0.13}{\percent}$, respectively (c.f.\ \cref{tab:summary-depolarizing}).
Regarding the rotated surface code, we compare our scheme with the recent results from Ref.~\cite{tan2023, higgott2023}, where corrections are applied after the decoding of $d$ rounds of syndrome measurements, achieving thresholds of $\SIrange{0.55}{0.94}{\percent}$, depending on the decoding technique used.~\footnote{We note that the thresholds claimed are for a single logical basis state different than $\ket{i}_L$, which in our simulations lowers the logical performance.}
Our implementation achieves a comparable pseudo-threshold (c.f.\ \cref{tab:summary-depolarizing}) by measuring half as many stabilizers.
Moreover, as comparing pseudo-threshold with threshold values can be misleading, we observe that our protocol achieves lower logical error rates at same physical rates.
A full comparison, which would require the use of similar decoding techniques, is beyond the scope of this paper.

\subsection{Noise model for neutral atoms}
\label{sec:biased-noise}

\begin{figure*}
    \centering
    \includegraphics[width=0.49\linewidth]{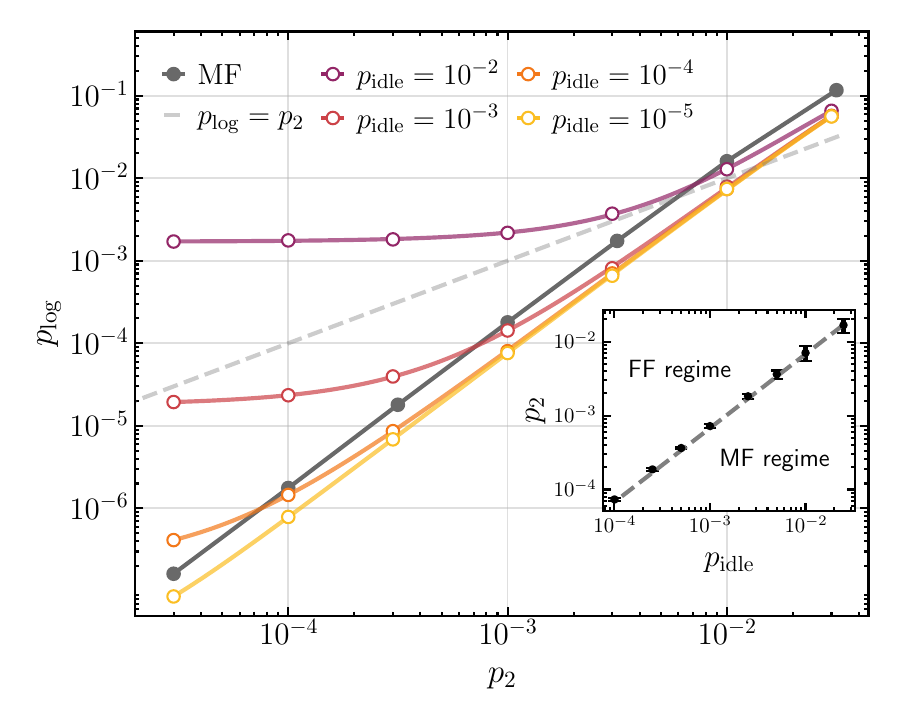}
    \begin{overpic}[width=0.49\linewidth]{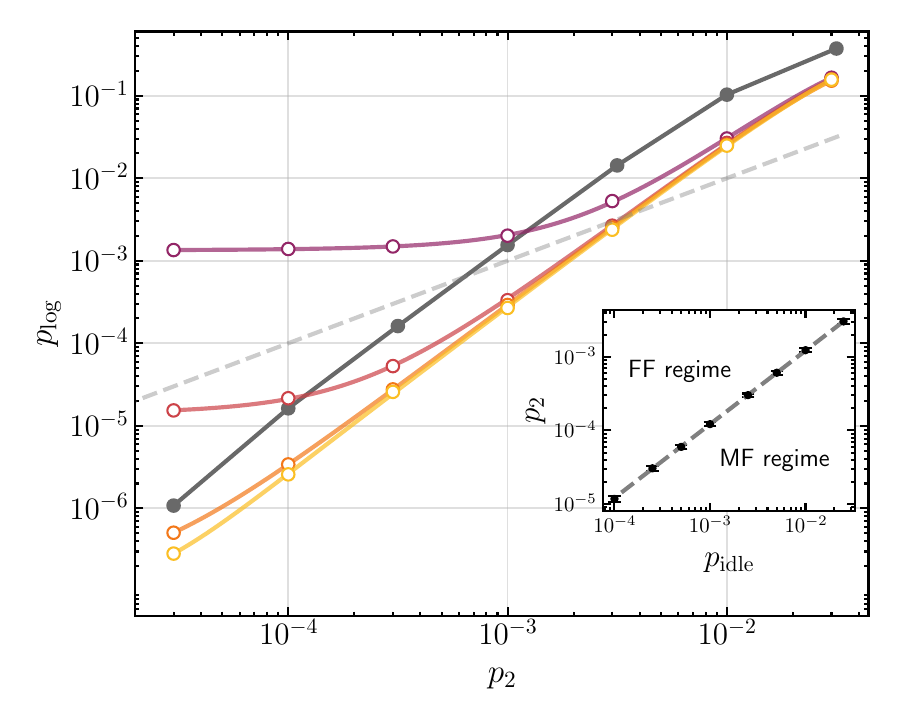}
        \put(-99, 77){(a)}
        \put(1, 77){(b)}
    \end{overpic}
    \vspace{-1.5em}
    \caption{
    Logical vs.\ physical error rate of our protocols for the (a) Bacon-Shor code, and (b) Steane's code,  under the simplified noise model for Rydberg-atom arrays described in \cref{sec:biased-noise}.
    The MF protocol is compared against FF implementations with varying idling error rates $p_{\mathrm{idle}}$.
    Datapoints of the FF protocol correspond to $15\,000$ simulations.
    First, we note there is a regime above a critical idling rate $\sim 10^{-2}$ where the logical error rate is always higher than the physical one.
    Secondly, for low gate error rates, idling errors dominate, and the MF approach, which under the described noise model is not affected, outperforms the FF protocol.
    The inset shows the regimes at which MF and FF implementations are advantageous as a function of $p_2$ and $p_\mathrm{idle}$.
    Current state-of-the-art experiments report $p_2 \approx 5 \times 10^{-3}$ across various atomic species~\cite{bluvstein2023, finkelstein2024, peper2024, radnaev2024}, while idling errors greatly vary between slightly less than $10^{-3}$ and $4\times 10^{-2}$~\cite{bluvstein2023, norcia2023, lis2023, graham2023}.
    Data points are numerical estimates of noise rates at which the protocols yield equivalent performances, matching the prediction in~\cref{eq:p2-MF-FF}.
    }
    \label{fig:plog_pidle_RN}
\end{figure*}

In neutral-atom architectures, the large difference between gate times and measurement times~\cite{shea2020, bluvstein2023} means that idling errors on data qubits \emph{during} measurements  cannot be neglected (cf.~\cref{app:noise-model:idling}).
Therefore, in this subsection, we compare the behavior of MF and FF protocols under a noise model which, albeit simplified, reproduces key features of neutral atom arrays.

For single-qubit gates, we consider \emph{quasi-static fluctuations} in the driving parameters.
In other words, we assume these fluctuations are slow enough compared to gate times, such that they can be considered static coherent errors for a single gate.
By averaging over independent realizations of such fluctuations, and performing the \emph{Pauli twirling approximation} (PTA) \cite{bennet1996, dur2005, emerson2007, silva2008, geller2013}, we obtain the following effective error channels for the $X$ and $H$ gates 
\begin{subequations}
    \begin{align}
    \mathcal{E}_{X} (\rho) &= \left( 1 - p_1 \right) \rho + \frac{p_1}{2} X \rho X + \frac{p_1}{2} Z \rho Z , \\
    \mathcal{E}_{H} (\rho) &= \left(1 - \frac{3 p_1}{4}\right) \rho
    + \frac{3 p_1}{8} X \rho X + \frac{3 p_1}{8} Z \rho Z.
    \end{align}
    \label{eq:channels-XH}
\end{subequations}
The details of the derivation are provided in~\cref{sec:noise-model}. 

The only native multi-qubit gates are $C^m Z$ gates. 
Their main source of infidelity is given by the decay from Rydberg states, which are auxiliary states with strong interatomic interactions used during the gate protocol to create entanglement between nearby atoms. 
By performing the PTA, this results in the noise channel (cf.\ \cref{sec:noise-model})
\begin{equation}
    \label{eq:channels-CnZ}
    \mathcal{E}_{C^{\ell-1}Z} (\rho) = \left(1 - p_\ell\right) \rho + \frac{p_\ell}{2^\ell - 1} \sum_{\mathclap{\substack{P \in \{I, Z\}^{\otimes \ell} \\ P \neq I^{\otimes \ell}}}} P \rho P^\dagger,
\end{equation}
i.e. controlled phase gates are only affected by $Z$-type errors.
For example, the Kraus operators for the $CZ$ gate are $\sqrt{p_2/3}\, Z_1$, $\sqrt{p_2/3}\, Z_2$, $\sqrt{p_2/3}\, Z_1 Z_2$.
We note that, under this biased noise, flags are superfluous and some constraints from \cref{sec:guidelines} could be relaxed.
Nevertheless, our noise model might be too simplistic; hence, for a fair comparison, we decided to run the generally FT circuits outlined in \cref{sec:codes}.

We simulate our implementations of the Bacon-Shor code in presence of such noise. 
We varied $p_2$, while fixing $p_1=p_2/5$ and $p_3 = 4p_2$. 
Furthermore, initializations and measurements were subject to bit-flips with respective rates $p_i = p_1/2$ and $p_m = p_2/2$. 
These values and ratios of error rates roughly describe recent experiments \cite{evered2023, bluvstein2023}.

We also simulate the FF protocols with varying idling error rate $p_{\mathrm{idle}}$, with the idling noise channel
\begin{equation}
    \mathcal{E}_{\mathrm{idle}}(\rho) = (1 - p_\mathrm{idle}) \rho + p_\mathrm{idle}\, Z \rho Z,
\end{equation}
affecting data qubits before corrections are applied.
Owing to the large difference between gate times and measurement times in neutral-atom arrays, we ignore idling errors in the MF setting.
In principle, one should take into account additional idling errors during shuttling operations to perform non-local gates, for both MF and FF protocols.
However, for the codes considered, the shuttling distances would be small on a 2D geometry, inducing a negligible time overhead compared to measurements.

Results are shown in~\cref{fig:plog_pidle_RN}.
Notably, the behavior of the logical error rate in FF protocols can be categorized into three regimes. 
For a protocol with $n$ data qubits and $N$ gates respectively, to leading order we have
\begin{equation}
    p_\mathrm{log}^\mathrm{FF} =
     \begin{cases}
       O(N^2\,p_2^2) &\; \mathrm{for}\; p_2 > p_\mathrm{idle} \\
       O(N\,n\,p_2\,p_\mathrm{idle}) &\; \mathrm{for}\; p_2 \lesssim p_\mathrm{idle} \\
       O(n^2\,p_\mathrm{idle}^2) &\; \mathrm{for}\; p_2 \ll p_\mathrm{idle} \\
     \end{cases}.
\end{equation}

Idling errors are independent of the quality of gates, thus, as the latter improves, idling errors may constitute an insurmountable limit in the performance of FF protocols.
In the first place, this implies that for some $p_\mathrm{idle}$ there is only a finite region where feed-forward gives an advantage over physical qubits. 
Secondly, as we considered idling errors negligible in the MF setting, the MF protocols eventually outperform the respective FF implementations as the quality of gates improves.
Considering that in FT MF protocols, the logical error rate scales as
\begin{equation}
    p_\mathrm{log}^\mathrm{MF} = O(N^2 \, p_2^2),
\end{equation}
the protocols yield equivalent performances when
\begin{equation}
    p_2^\mathrm{MF=FF} = O\!\left(\frac{n}{N}\right) p_\mathrm{idle}. \label{eq:p2-MF-FF}
\end{equation}
This matches the result in the inset of~\cref{fig:plog_pidle_RN}. 
We estimate that for the Bacon-Shor and Steane's protocols, the proportionalities are $\sim 0.68$ and $\sim 0.12$, respectively, in both cases roughly $4 n / N$. 

Finally, we note that current state-of-the-art experiments with neutral atoms have achieved $p_2 \approx 5 \times 10^{-3}$ and $p_\mathrm{idle} \lesssim 10^{-3}$~\cite{bluvstein2023}.
Such estimates place this experiment in the FF regime, albeit not far from the MF.
Additional results without idling errors can be found in \cref{app:additional-numerics}.
For a practical implementation, the choice of FF vs.\ MF may not be dictated only by the logical error rate, but also by considerations on the \emph{QEC cycle rate}.
Considering the large difference between measurement times and gate times, MF QEC can be a preferable setting for performing repeated QEC cycles, even if this incurs in a small performance penalty.
Additionally, in neutral atoms long idling times incur in atom loss, which are outside of the computational subspace.
Such difficulties have, to date, prevented the practical demonstration of repeated QEC cycles with neutral atoms.

\section{Summary and outlook}
\label{sec:conclusion}

In summary, we have proposed several implementations of measurement-free QEC schemes based on syndrome redundancy and single-shot flags.
The classical logic traditionally involved in QEC schemes is moved \emph{within} the quantum circuit and is performed using quantum gates.
These implementations have been designed with near-term quantum hardware in mind, with focus on the smallest non-trivial CSS codes.
By using the guiding principles outlined in \cref{sec:guidelines}, we design quantum circuits which are FT, with lower overhead than in alternative proposals~\cite{heussen2023a, perlin2023}.

The measurement-free QEC protocols for the Bacon-Shor code and Shor's code provide examples that we constructed using such heuristics, as discussed in detail in \cref{sec:codes}.
We furthermore show that these guidelines can be extended to any CSS code, although one needs to adapt the correction step to remain FT.
This can be achieved by using auxiliary qubits for decomposing larger multi-controlled operations, as is discussed in the cases of the surface code and Steane's code in \cref{app:surface-code,app:steane-code}.
By careful circuit design, the additional overhead of measurement-free QEC can be kept to a minimum.

The development of such measurement-free techniques could be beneficial for the design of more traditional feed-forward implementations as well.
The proposed redundant syndrome extraction is of interest for so-called single-shot decoding~\cite{bombin2015}, which allows for a robust protocol without repeating extraction rounds.
The overhead of repeating syndrome extraction or using flags could be reduced, simply by extracting an overcomplete set of stabilizers. 

We have benchmarked our scheme with numerical simulations under depolarizing noise, both for the measurement-free and feed-forward implementations, as summarized in~\cref{tab:summary-depolarizing}. 
The feed-forward protocols achieve performances comparable with state-of-the-art techniques, thus they represent a competitive single-shot alternative. 
Our measurement-free protocols have pseudo-thresholds that are only about a factor of 2 smaller than the ones of the comparable feed-forward implementations.
Particularly interesting are the ones that can correct the most higher-weight errors and use the fewest gates, with the Bacon-Shor code showing the best performance overall.

Measurement-free schemes are particularly interesting in quantum-computing platforms where measurements are a main bottleneck, such as neutral-atom arrays.
For such platforms, we have derived a simplified noise model, which leads to improved pseudo-thresholds---within the reach of current state-of-the-art experimental error rates~\cite{bluvstein2023}.
Additionally, we have performed a more realistic comparison with feed-forward QEC by adding idling errors during measurements, and observed that, as the quality of the gates improves, idling errors significantly drag down the performance of the protocols. 
In this context, measurement-free implementations, where idling errors play less of a role or even can become negligible, outperform the respective feed-forward protocols.

Neutral-atom arrays have additional features particularly suited for measurement-free QEC.
Multi-controlled gates---such as the $CCZ$, necessary for the correction step---are natively supported on the hardware.
Additionally, shuttling via optical tweezers could enable the increased connectivity required.
We emphasize that these numerical results should be taken not too literally: the noise model for Rydberg atoms is undoubtedly overly simplistic, as it neglects other sources of errors, such as atom loss or leakage outside of the computational subspace.
Including these effects into a refined noise model would increase the numerical cost, and more importantly, will require further assumptions on the specific qubit encoding used.
We leave this as a possible direction for future work.

We emphasize that the proposed new measurement-free error correction schemes are not solely relevant to neutral-atom arrays.
The only requirement is the availability of fast and locally addressable reset operations, or, alternatively, of a sufficiently large reservoir of fresh ancillae.
In this case, a full or partial measurement-free scheme can be beneficial, as one can trade measurements for quantum gates.
This could lead to a faster or less error-prone QEC implementation, depending on the hardware specifics.
With Rydberg atom arrays in mind, we focused on the case of native multi-controlled gates, such as $CCZ$s.
As shown in Ref.~\cite{heussen2023a}, such gates can be fault-tolerantly decomposed rather efficiently if the controls are on ancillas~\footnote{%
    A single Toffoli gate with resets on the controls decomposes into four CNOTs and six single-qubit gates.
}.
We therefore expect that our correction circuits can be adapted to use only one-qubit and two-qubit gates, with a moderate performance overhead.

Having near-term devices in mind, we have only considered $d=3$ codes.
Scaling up error-correcting capabilities is hard to achieve in a purely measurement-free setting.
For higher-distance codes, the decoding logic becomes non-trivial and, accordingly, it becomes much more challenging to design a FT correction circuit.
Code concatenation could provide an alternative approach for scaling to larger distances.
The implementations presented in this paper can be realized with a native gate set of $\{H, X, CZ, CCZ\}$.
If this set can be implemented at the logical level, then the code could be concatenated with itself.
This could possibly be achieved with Steane's code, as it supports all Clifford gates as well as the so-called pieceable FT $CCZ$ gates~\cite{yoder2016}.
Similarly, concatenation could be achieved with codes supporting fold-transversal gates such as the surface code or the Bacon-Shor code.
Alternatively, rather than attempting to scale up measurement-free QEC, we believe that small measurement-free blocks could be beneficially concatenated within conventional feed-forward protocols.
In this setting, measurements could be performed much less frequently, while the measurement-free units run fast and autonomously.
Recent experiments have demonstrated that neutral-atom arrays are well-suited to implementing small error-correcting blocks in a modular fashion \cite{bluvstein2023}.
By avoiding or minimizing measurements, the measurement-free approach could lead to improvements in the logical qubit fidelity or, at least, provide significant speedups of the durations of QEC cycles.

Concatenation also opens up the possibility of implementing universal quantum computation.
Steane's code could be concatenated with a code with transversal $T$ gates, such as three-dimensional (3D) color codes or triorthogonal codes~\cite{paetznick2013,jochym-oconnor2014,chamberland2017}.
In such constructions, code distance is traded for a larger set of transversal gates.
Incorporating measurement-free QEC into such logical gate constructions could be key to achieving fast universal quantum computation.

\begin{acknowledgments}
We acknowledge inspiring discussions with Alexandru Paler and Alexander Glätzle.
We would especially like to thank Johannes Zeiher for fruitful discussions and insightful feedback on this manuscript.
This research is part of the Munich Quantum Valley (K-8), which is supported by the Bavarian state government with funds from the Hightech Agenda Bayern Plus.
M.M. additionally acknowledges support by the BMBF project MUNIQC-ATOMS (Grant No.~13N16070), by the Deutsche Forschungsgemeinschaft (DFG, German Research Foundation) under Germany’s Excellence Strategy ‘Cluster of Excellence Matter and Light for Quantum Computing (ML4Q) EXC 2004/1’ 390534769, 
and from the European Union’s Horizon Europe research and innovation programme under Grant Agreement No. 101114305 (“MILLENION-SGA1” EU Project) and ERC Starting Grant QNets through Grant No.~804247.

\end{acknowledgments}

\appendix
\section{Requirements on redundant syndromes}
\label{appendix:redundant permut}

This appendix defines a constraint on the order in which stabilizers are extracted within our implementations of QEC protocols for $d = 3$ codes.  
Error syndromes obtained from the extraction of $m$ stabilizers are bit-strings \mbox{$\mathbf{b} = (x_1,\, \dots,\, x_m)$}.
Ideal decoding associates syndromes to correction operations.
Let us call a syndrome bit-string $\Tilde{\mathbf{b}}_i$ \emph{active} if it leads to a correction operation different than the identity. 
Since the extraction of stabilizer eigenvalues is sequential, the order matters.
In particular, we require that, given the set of active syndromes $\Tilde{\mathcal{S}}=\{\Tilde{\mathbf{b}}_i\}$
\begin{equation}
    \Tilde{\mathbf{b}}_i \neq R_s \big(\Tilde{\mathbf{b}}_j\big) \quad \forall i,j,s,
    \label{eq:redundancy-requirement}
\end{equation}
where $R_s \big(\Tilde{\mathbf{b}}_j\big)$ returns $\Tilde{\mathbf{b}}_j$ with the first $s$ 1s turned into 0s.

\begin{figure}[hb]
    \includegraphics[width=0.95\linewidth]{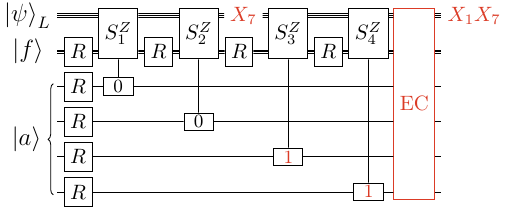}
    \vspace{-1.25em}
    \caption{%
        Circuit for detecting bit-flips in the Steane code, illustrating the requirement for two extra syndromes to be FT.
        $S^Z_1$, $S^Z_2$, $S^Z_3$, defined in \cref{eq:steane stabilizers}, are the $Z$-type stabilizers generators.
        $S^Z_4$ is the product of the three. 
        All of them involve data qubit $7$. 
        An error $X_7$ occurring at the end of the $S_2^Z$ syndrome is measured by the following stabilizers.
        This returns the syndrome $0011$, leading to the correction $X_1$. 
        Together with the previous error $X_7$, this forms an uncorrectable weight-two error.
        We have exhaustively verified that no ordering of any combination of four stabilizers can satisfy \cref{eq:redundancy-requirement}.
        An appropriate redundancy is instead used in~\cref{fig:SteaneCircs}.
    }
    \label{fig:faultySteane}
\end{figure}

This requirement stems from the fact that stabilizer extractions are prone to errors. 
Assume a logical codeword is passed as input to the QEC circuit, and that we are measuring $Z$-type stabilizers. 
Assume all stabilizer extractions are fault-free except for a single one, the $s^\mathrm{th}$ involving the $i^\mathrm{th}$ data qubit. 
During this extraction, a bit-flip $X_i$ occurs on the said data qubit. 
The syndrome bit-string $\Tilde{\mathbf{b}}_i$, leading to the correction $X_i$, is not observed, as the data were fault-free in the first $s$ extractions involving data qubit $i$. 
Instead, $\Tilde{\mathbf{b}}_j = R_s \big(\tilde{\mathbf{b}}_i\big)$ is observed. 
If $\Tilde{\mathbf{b}}_j$ leads to correcting a different data qubit, the protocol accumulates two different errors. 
This would spoil fault-tolerance.
Analogous considerations hold for phase flips occurring during extraction of $X$-type stabilizers.
This is illustrated in \cref{fig:faultySteane} for the Steane code---further described in \cref{app:steane-code}.

The requirement in \cref{eq:redundancy-requirement} can be generalized to non-CSS codes by adding additional constraints, arising from the fact that different operators are measured on the same qubits during the syndrome extraction procedure. 
Thus, we expect our strategy to be applicable for non-CSS codes, albeit with higher overhead.

\section{Additional CSS codes}
\label{appendix:surf-steane}
This appendix contains the MF implementations for the rotated surface code and Steane's code.

\subsection{Surface code}
\label{app:surface-code}

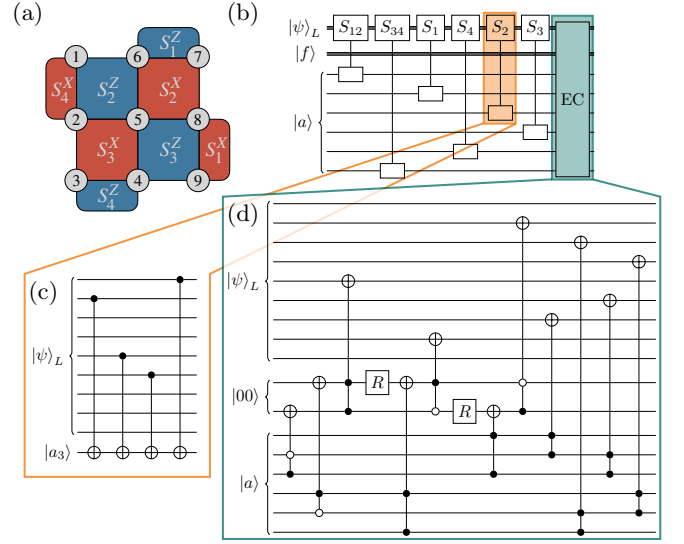
\begin{figure}[!t]
    \vspace{1em}
    \begin{overpic}[width=\linewidth]{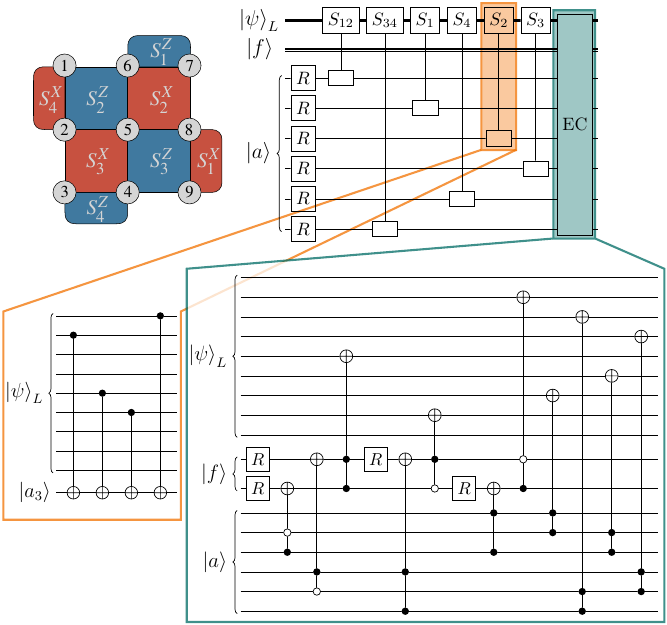}
        \put(1, 90){(a)}
        \put(30, 90){(b)}
        \put(1, 43){(c)}
        \put(29, 49){(d)}
    \end{overpic}
    \phantomsubfloat{\label{fig:surf-circ-stabs}}
    \phantomsubfloat{\label{fig:surf-circ-circ}}
    \phantomsubfloat{\label{fig:surf-circ-extr}}
    \phantomsubfloat{\label{fig:surf-circ-corr}}
    \vspace{-0.25em}
    \caption{%
        Fault-tolerant MF implementation for the surface code, using a total of 17 qubits (9 data qubits, 6 ancilla and 2 intermediary qubits).
        (a) Stabilizers of the $d = 3$ surface code.
        (b) General scheme for single-shot correction of a bit- or phase-flip. 
        This involves a redundant extraction of stabilizers, and an error correction ($\mathrm{EC}$) block.
        (c) Syndrome measurement of the $S_2^Z$ stabilizer. 
        As for the other stabilizers, it does not require any flags, as long as the first two and the last two extracted operators constitute a weight-two correctable error, which is allowed to propagate onto the data qubit register.
        (d)  FT MF correction circuit for bit-flips. 
        The circuit for correcting phase-flips is analogous, with $CCZ$ gates substituting the Toffoli gates acting on data qubits. 
    }
    \label{fig:SurfCircs}
\end{figure}

The smallest error-correcting surface code $\llbracket 9, 1, 3 \rrbracket$ \cite{bravyi1998, dennis2002, fowler2012} has the following stabilizer generators
\begin{equation}
    \begin{aligned}[c]
    S_1^X &= X_8 X_9, \\
    S_2^X &= X_5 X_6 X_7 X_8, \\
    S_3^X &= X_2 X_3 X_4 X_5, \\
    S_4^X &= X_1 X_2,
    \end{aligned}
    \qquad \qquad
    \begin{aligned}[c]
    S_1^Z &= Z_6 Z_7, \\
    S_2^Z &= Z_1 Z_2 Z_5 Z_6, \\
    S_3^Z &= Z_4 Z_5 Z_8 Z_9, \\
    S_4^Z &= Z_3 Z_4,
    \label{eq:surface stabilizers}
    \end{aligned}
\end{equation}
as shown in \cref{fig:surf-circ-stabs}.
Its logical operators can be chosen to be $X_L = X_1 X_6 X_7$ and $Z_L = Z_1 Z_2 Z_3$.
A FT MF codeword encoding has been introduced in \cite{goto2023}.
\cref{fig:SurfCircs} illustrates our FT MF design with a total of $17$ qubits. 

We notice that the code has a certain symmetry. 
Stabilizers come in pairs that are specular with respect to qubit $5$, and so is the optimal decoding, which is shown in \cref{table:surf1}.
Therefore, we added two new syndromes, $S_{12} = S_1 S_2$ and $S_{34} = S_3 S_4$, that preserve the symmetry. 

The code is very modular, as stabilizers are well-localized and share only few qubits. 
This makes the code capable of correcting multiple weight-two errors: all the pairs $X_i X_j$ such that there exists no $k$ for which $X_L = X_i X_j X_k$, and similarly for $Z$. 
As noted in Ref.~\cite{tomita2014}, this implies that the extraction of weight-four stabilizers does not require flags if gates are opportunely ordered, as in \cref{fig:surf-circ-extr}. 
Trivially, the read-out of weight-two stabilizers does not require flags either. 

The symmetry of the code allows for a simple design of a MF correction circuit, capable of implementing the optimal decoding of \cref{table:surf1}. 
The correction circuit is illustrated in \cref{fig:surf-circ-corr}. 
Note how it uses an intermediate register between ancillae and data qubits. 
The intermediary register prevents single faults to cause errors on both ancillae and data qubits, in a fashion similar to \cite{heussen2023a}, but with fewer qubits, as well as gates with support on fewer qubits (Toffoli-type gates, supported on three qubits).

\begin{table}[h!b]
    \def\arraystretch{1.05}
    \begin{tabular}{c|c|c}
        \hline\hline
        \makecell{Syndrome \\ ($S_1, S_2, S_3, S_4$)} & $X$-Correction & $Z$-Correction \\
        \hline

        01($xy\neq$10) & $X_2$ & $Z_6$ \\
        $xy$01 & $X_3$ & $Z_1$ \\
        $xy$11 & $X_4$ & $Z_2$\\

        0110 & $X_5$ & $Z_5$\\

        11$xy$ & $X_6$ & $Z_8$\\
        10$xy$ & $X_7$ & $Z_9$\\
        ($xy\neq$01)10 & $X_8$ & $Z_4$ \\

        \hline\hline
    \end{tabular}
    \caption{Optimal one-round decoding for the surface code.
    $xy$ denote digits of the syndrome that are irrelevant for that specific correction.
    This enables weight-two corrections: for example bit-flip syndrome $1101$ leads to correction $X_3 X_6$. 
    \label{table:surf1}
    \vspace{-1.5em}
    }
\end{table}

\subsection{Steane's code}
\label{app:steane-code}

\begin{figure}[t]
    \vspace{1em}
    \begin{overpic}[width=\linewidth]{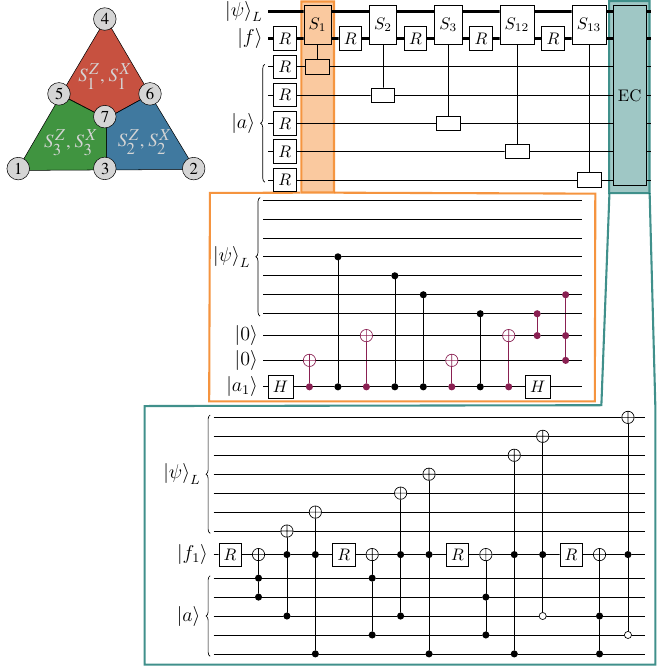}
        \put(0, 97){(a)}
        \put(26, 97){(b)}
        \put(32, 67){(c)}
        \put(23, 35){(d)}
    \end{overpic}
    \phantomsubfloat{\label{fig:steane-circ-stabs}}
    \phantomsubfloat{\label{fig:steane-circ-circ}}
    \phantomsubfloat{\label{fig:steane-circ-extr}}
    \phantomsubfloat{\label{fig:steane-circ-corr}}
    \vspace{-0.25em}
    \caption{%
        Fault-tolerant MF implementation of the Steane code, using 14 qubits. 
        (a) Stabilizers of Steane's code, the smallest color code~\cite{bombin2006}. 
        (b)  General scheme for single-shot correction of a bit- or phase-flip.
        This involves a redundant flagged extraction of stabilizers, and an error correction ($\mathrm{EC}$) block.
        (c) Syndrome measurement of the $S_1^Z$ stabilizer.
        Single-shot flags, whose operations are highlighted in purple, prevent uncorrectable errors from spreading, in the same fashion as depicted in~\cref{fig:shor}. 
        (d) Correction circuit for bit-flips. The correction of phase-flips uses an analogous circuit, with $CCZ$ gates replacing the Toffoli gates acting on data qubits. 
    }
    \label{fig:SteaneCircs}
    \vspace{-0.15em}
\end{figure}
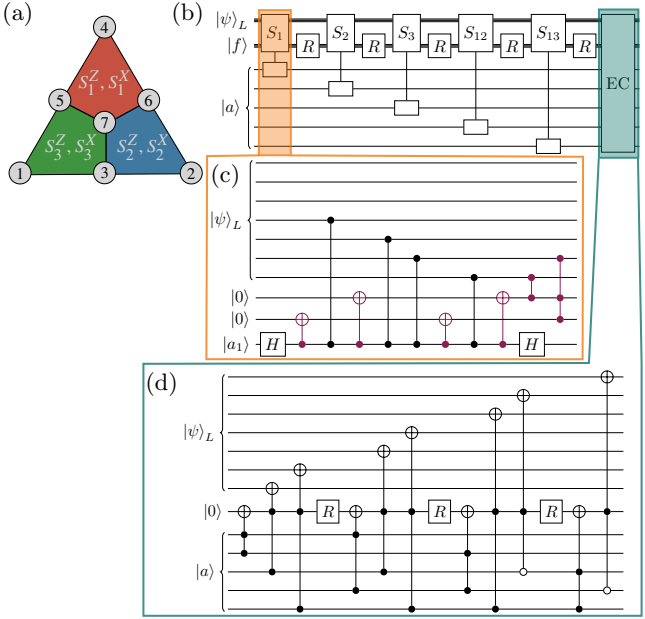

Steane's code $\llbracket 7, 1, 3 \rrbracket$ \cite{steane1996} has the following stabilizer generators
\begin{equation}
    \begin{aligned}[c]
    S_1^X = X_4 X_5 X_6 X_7, \\
    S_2^X = X_2 X_3 X_6 X_7, \\
    S_3^X = X_1 X_3 X_5 X_7,
    \end{aligned}
    \qquad \qquad
    \begin{aligned}[c]
    S_1^Z = Z_4 Z_5 Z_6 Z_7, \\
    S_2^Z = Z_2 Z_3 Z_6 Z_7, \\
    S_3^Z = Z_1 Z_3 Z_5 Z_7, \label{eq:steane stabilizers}
    \end{aligned}
\end{equation}
as illustrated in \cref{fig:steane-circ-stabs}.
Its logical operators can be chosen to be $X_L = X_1 X_2 X_3$ and $Z_L = Z_1 Z_2 Z_3$.
The Steane code has the computational advantage of providing transversal Clifford logical operations.
A FT MF encoding of the logical states has been recently introduced in Ref.~\cite{heussen2023a}.
\Cref{fig:SteaneCircs} illustrates our FT MF 14-qubits design. 

As shown, a single extraction block (for bit- or phase-flips) measures five stabilizers, two more than the minimum required. 
Besides those in \cref{eq:steane stabilizers}, we measure $S_{12} = S_1 S_2$, and $S_{13}=S_2 S_3$. 
In fact, by exhaustively checking the possible combinations and permutations, it can be verified that the addition of only one stabilizer is not sufficient to satisfy the constraint outlined in \cref{appendix:redundant permut}, and therefore fault-tolerance cannot be achieved. 
A full redundancy, achievable by measuring seven syndromes as in Refs.~\cite{crow2016, perlin2023}, is not only unnecessary, but perhaps even detrimental as it would largely increase both the number of qubits and the number of gates employed.

As shown for stabilizer $S^Z_1$ in \cref{fig:steane-circ-extr}, extractions are supported by single-shot flags, as for our implementation of Shor's code in~\cref{sec:shor}. 
In contrast to conventional flagged circuits \cite{chao2018}, this design does not require measurements, as the same extraction circuit is run independently of the values of the flags. 

Finally, \cref{fig:steane-circ-corr} outlines our FT MF correction block. 
As for the surface code, it uses an intermediate register, here a single qubit, between ancillae and data qubits to avoid single faults to spread. 

\section{Simplified noise model for neutral atoms}
\label{sec:noise-model}
In this Appendix, we construct a simple effective noise model to account for imperfections in neutral-atom arrays.

\subsection{Single-qubit gates}

We begin by considering a laser-driven two-level system.
The Hamiltonian between the two computational states reads as
\begin{align}
    \mathcal{H} &= \frac{\Omega}{2}\left( e^{i \phi} \ket{1}\bra{0} + \mathrm{H.c.} \right) + \frac{\Delta}{2}\left(\ket{0}\bra{0} - \ket{1}\bra{1}\right) \nonumber\\
    &= \frac{\Omega}{2}\left(\cos\phi X + \sin\phi Y \right) + \frac{\Delta}{2} Z,
\end{align}
where $\Omega$ is the Rabi frequency, $\phi$ is the laser phase and $\Delta$ the detuning of the laser with respect to the transition frequency.
By evolving the system for a time $t$ (assuming, for simplicity, a constant profile for $\Omega$ and $\Delta$), we obtain the unitary $U = \exp \left({-i \mathcal{H} t} \right)$
\begin{equation}
    \label{eq:unitary-single-qubit}
    U = \cos\theta - i \sin\theta \left( \frac{\Omega}{\tilde{\Omega}} \cos\phi X + \frac{\Omega}{\tilde{\Omega}} \sin\phi Y + \frac{\Delta}{\tilde{\Omega}} Z\right),
\end{equation}
where $\tilde{\Omega} = \sqrt{\Omega^2 + \Delta^2}$ and $\theta = \frac{\tilde{\Omega}t}{2}$.
In the following, we consider \emph{static} fluctuations of the driving parameters, which is physically motivated when these fluctuations happen over time scales much larger than the typical gate times.
We therefore average \cref{eq:unitary-single-qubit} over different realizations with the driving parameters drawn from a Gaussian distribution.
This generates the channel $\Lambda(\rho) = \overline{U \rho U^\dagger}$, where the bar denotes the averaging over the stochastic variable.

For example, we account for phase noise by setting $\phi = \xi$, where $\xi \sim \mathcal{N}(0, \sigma^2)$.
This corresponds to slow fluctuations of the laser phase during the driving.
It also corresponds to a simplified model of the momentum transfer between photons and the atom in the laser field, which imprints a phase factor $e^{i \mathbf{k} \cdot \mathbf{x}}$ for a wavevector $\mathbf{k}$ and the atomic center of mass $\mathbf{x}$.

Considering the noiseless gate $U_0$ generated by \cref{eq:unitary-single-qubit} at $\phi = 0$, we can then perform the average over $\xi$, and obtain the channel
\begin{widetext}
\begin{equation}
    \label{eq:channel-phase-general}
    \Lambda^{\mathrm{ph}}(\rho) \simeq 
    \left(1 - \frac{\sigma^2}{2}\right) \! U_0 \rho U_0^\dagger +
    \frac{\sigma^2}{2} \cos^2{\theta}\rho -
    i \frac{\sigma^2}{2} \cos{\theta} \sin{\theta} \frac{\Delta}{\tilde{\Omega}} \left(Z \rho - \rho Z  \right)
    + \sigma^2 \sin^2{\theta} 
    \left[
    - \frac{1}{2}\frac{\Omega^2}{\Tilde{\Omega}^2} X\rho X
    + \frac{\Omega^2}{\Tilde{\Omega}^2} Y\rho Y
    + \frac{1}{2} \frac{\Delta^2}{\Tilde{\Omega}^2} Z\rho Z
    \right].
\end{equation}
where we have applied the following first-order approximations
\begin{equation*}
    \overline{e^{i \phi}} = \overline{\cos{\phi}}= e^{-\sigma^2\!/2} \simeq 1 - \frac{\sigma^2}{2} , \qquad 
    \overline{\cos^2\phi} \simeq 1 - \sigma^2 , \qquad
    \overline{\sin\phi \cos\phi} \simeq 0 , \qquad
    \overline{\sin^2\phi} \simeq \sigma^2 .
\end{equation*}
\end{widetext}

For the case of the $X$ gate ($\Delta=0$ and $t=\pi/\Omega$) and  $H$ gate ($\Delta=\Omega$ and $t=\pi/\tilde{\Omega}$), \cref{eq:channel-phase-general} reduces to
\begin{subequations}
    \label{eq:channel-phase}
    \begin{align}
    \Lambda_{X}^{\mathrm{ph}}(\rho) =&
    \left(1 - \sigma^2\right) X \rho X + \sigma^2 Y \rho Y , \\
    \Lambda_{H}^{\mathrm{ph}}(\rho) =&
    \left(1 - \frac{\sigma^2}{2}\right) H \rho H \nonumber \\& -
    \frac{\sigma^2}{4} X \rho X +
    \frac{\sigma^2}{2} Y \rho Y +
    \frac{\sigma^2}{4} Z \rho Z .
    \end{align}
\end{subequations}
In the context of quantum error correction, we typically express the noise operation as a channel \emph{after} the desired gate $U_0$, i.e.
\begin{equation}
    \Lambda_{U_0}(\rho) =  \mathcal{E}_{U_0}\!\left(U_0 \rho U_0^\dagger\right),
\end{equation}
For the phase noise in \cref{eq:channel-phase}, we then obtain
\begin{subequations}
    \begin{align}
    \mathcal{E}_{X}^{\mathrm{ph}}(\rho) =& \left(1 - p_{\mathrm{ph}}\right) \rho + p_{\mathrm{ph}} Z \rho Z , \\
    \mathcal{E}_{H}^{\mathrm{ph}}(\rho) =&
    \left(1 - \frac{p_{\mathrm{ph}}}{2}\right) \rho -
    \frac{p_{\mathrm{ph}}}{4} XH \rho HX \nonumber \\&+
    \frac{p_{\mathrm{ph}}}{2} YH \rho HY +
    \frac{p_{\mathrm{ph}}}{4} ZH \rho HZ,
    \end{align}
\end{subequations}
where we define $p_{\mathrm{ph}} = \sigma^2$.

We now perform the \emph{Pauli twirling approximation} (PTA) \begin{equation}
    \label{eq:pauli-twirling}
    \tilde{\mathcal{E}}(\rho) = \sum_{P \in \mathcal{P}_\ell}  P^\dagger \mathcal{E}\!\left(P \rho P^\dagger\right) P,
\end{equation}
to approximate the channels above with the closest possible Pauli channel \cite{bennet1996, dur2005, emerson2007, silva2008, geller2013}.
This can be experimentally realized by physically applying a random Pauli string before and after the gate.

Using the PTA in \cref{eq:pauli-twirling}, the latter becomes
\begin{equation}
    \tilde{\mathcal{E}}_{H}^{\mathrm{ph}}(\rho) = \left(1 - \frac{p_{\mathrm{ph}}}{2}\right) \rho + \frac{p_{\mathrm{ph}}}{4} \left( X \rho X + Z \rho Z \right).
\end{equation}

Similarly, we can account for fluctuations of the pulse area by setting $\theta =\theta_0 (1 + \xi)$ in \cref{eq:unitary-single-qubit} where $\xi \sim \mathcal{N}(0, \sigma^2)$.
This can be seen as an amplitude fluctuation, and fluctuations on the amplitude can, in the case of Raman transitions,  induce correlated fluctuations on the detuning~\cite{day2022}.
By performing similar steps to the case of phase noise, we obtain
\begin{equation}
    \label{eq:channel-time-general}
    \Lambda^{\mathrm{time}}(\rho) \simeq 
    (1-2\theta_0^2\sigma^2) U_0 \rho U_0^\dagger 
    + \theta_0^2 \sigma^2 (\rho + K\rho K),
\end{equation}
where $K = \frac{\Omega}{\tilde{\Omega}} X + \frac{\Delta}{\tilde{\Omega}} Z$.
For the $X$ and $H$ gates this corresponds to
\begin{subequations}
    \label{eq:channel-time}
    \begin{align}
    \Lambda_{X}^{\mathrm{time}}(\rho) =&
    \left(1 - \frac{\pi^2\sigma^2}{4}\right) X \rho X + \frac{\pi^2\sigma^2}{4} \rho , \\
    \Lambda_{H}^{\mathrm{time}}(\rho) =&
    \left(1 - \frac{\pi^2\sigma^2}{4}\right) H \rho H + \frac{\pi^2\sigma^2}{4} \rho .
    \end{align}
\end{subequations}
This leads to
\begin{subequations}
    \begin{align}
    \mathcal{E}_{X}^{\mathrm{time}}(\rho) =& \left(1 - p_{\mathrm{time}}\right) \rho + p_{\mathrm{time}} X \rho X , \\
    \mathcal{E}_{H}^{\mathrm{time}}(\rho) =& \left(1 - p_{\mathrm{time}}\right) \rho + p_{\mathrm{time}} H \rho H ,
    \end{align}
\end{subequations}
where we define $p_{\mathrm{time}} = \pi^2\sigma^2/4$.
Performing the PTA on the error channel for the $H$ gate yields 
\begin{equation}
    \tilde{\mathcal{E}}_{H}^{\mathrm{time}}(\rho) = \left(1 - p_{\mathrm{time}}\right) \rho + \frac{p_{\mathrm{time}}}{2} \left( X \rho X + Z \rho Z \right).
\end{equation}

Putting everything together, choosing $p_{\mathrm{ph}} = p_{\mathrm{time}}$ yields the result in the main text, cf.~\cref{eq:channels-XH}, with $p_1 = p_{\mathrm{ph}} + p_{\mathrm{time}}$.

\begin{figure*}[t]
    \vspace{1em}
    \begin{overpic}[width=\textwidth]{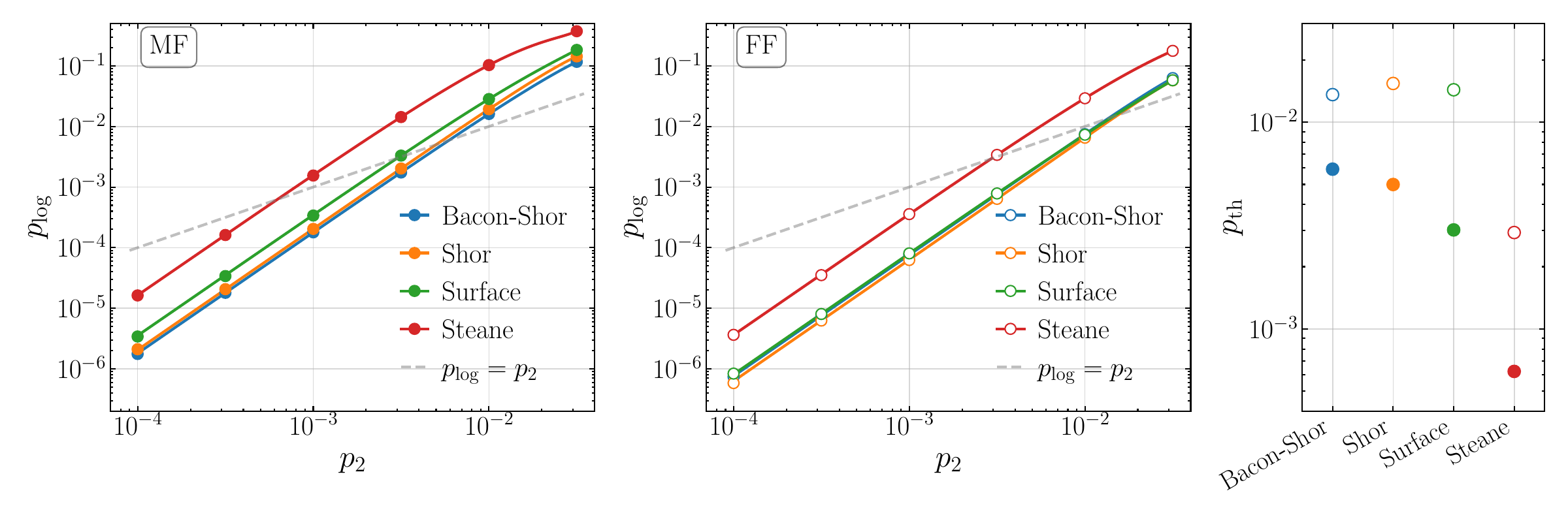}
        \put(0, 32){(a)}
        \put(40, 32){(b)}
        \put(78, 32){(c)}
    \end{overpic}
    \phantomsubfloat{\label{fig:RDN_MF}}
    \phantomsubfloat{\label{fig:RDN_FF}}
    \phantomsubfloat{\label{fig:RDN_pth}}
    \vspace{-2.5em}
    \caption{
        Logical qubit error rate vs.\ physical one, under the simplified noise model for neutral atoms described in \cref{sec:biased-noise}, for the (a) measurement-free and (b) feed-forward implementation.
        Solid lines correspond to the numerical fits used to estimate the pseudo-thresholds, similar to \cref{fig:plogs_FDN}.
        (c) Numerical estimates of the pseudo-thresholds.
    }
    \label{fig:plogs_RDN}
\end{figure*}

\subsection{Multi-controlled gates}
An entangling gate can be realized in neutral atoms by driving the $\ket{1}$ state to an auxiliary Rydberg $\ket{r}$ state.
Ideally, the $\ket{r}$ state is completely unoccupied after the protocol has been performed.
The main source of error comes from the spontaneous decay from the $\ket{r}$ state.
Here we assume simplistically that the decay from the $\ket{r}$ state can only happen to the $\ket{1}$ state.
Such effects have been explored to mitigate errors arising from leakage outside of the computational subspace, and are broadly referred to as \emph{erasure conversion}, both in theoretical proposals~\cite{cong2022,sahay2023}, as well as in experimental setups~\cite{wu2022,ma2023}.
Here we follow a similar approach to Ref.~\cite{sahay2023}.
We assume that with perfect erasure conversion, all remaining populations in the Rydberg state (for example due to driving imperfections) are rought back to the $\ket{1}$ state.
Importantly, we assume that the erasure conversion can be achieved solely by incoherent repumping, and not via measurements.
We then derive the associated quantum channel.

A $C^m Z$ gate is supported on $\ell = m + 1$ qubits.
The error channel is $\mathcal{E}_{C^m Z}^{\mathrm{decay}}(\rho) = \sum_\mathbf{s} K_\mathbf{s} \rho K_\mathbf{s}^\dagger$, where the Kraus operators are indexed using the basis state $\ket{\mathbf{s}} = \ket{s_1, s_2, \dots, s_\ell}$ that is subject to decay, 
\begin{equation}
    K_\mathbf{s} = 
    \begin{cases}
        \displaystyle \ket{\mathbf{0}} \bra{\mathbf{0}} + \sqrt{1 - p} \sum_{\mathbf{s'} \neq \mathbf{0}} \ket{\mathbf{s'}} \bra{\mathbf{s'}} & \mbox{if}~\mathbf{s} = \mathbf{0}, \\
        \displaystyle \sqrt{p} \ket{\mathbf{s}} \bra{\mathbf{s}} & \mbox{otherwise},
    \end{cases}
\end{equation}
where we define $\ket{\mathbf{0}} = \ket{0 0 \dots 0 0}$.
We assume that all the states $\mathbf{s} \neq \mathbf{0}$ are subject to the same decay probability $p$ during the gate protocol.
This is realistic, since in the blockaded regime only one $\ket{r}$ state can be excited at a given time, so their decay probability should be similar.
Expanding to first order in $p$ we obtain
\begin{equation}
    K_\mathbf{s} \rho K_\mathbf{s}^\dagger \simeq
    \begin{cases}
        \displaystyle (1 - p) \rho + \frac{p}{2} \left(\ket{\mathbf{0}} \bra{\mathbf{0}} \rho + \rho \ket{\mathbf{0}} \bra{\mathbf{0}} \right) & \mbox{if}~\mathbf{s} = \mathbf{0} , \\
        \displaystyle \; p \ket{\mathbf{s}} \bra{\mathbf{s}} \rho \ket{\mathbf{s}} \bra{\mathbf{s}} & \mbox{otherwise} .
    \end{cases}
\end{equation}
Using the identities $\ket{0} \bra{0} = \frac{1}{2}(I + Z)$ and $\ket{1} \bra{1} = \frac{1}{2}(I - Z)$ we twirl each term with \cref{eq:pauli-twirling} to obtain
\begin{equation}
    \widetilde{K_\mathbf{s} \rho K_\mathbf{s}^\dagger} \simeq
    \begin{cases}
        \displaystyle \left[1 - \left(1 - \frac{1}{2^{\ell}} \right)p\right] \rho & \mbox{if}~\mathbf{s} = \mathbf{0} , \\
        \displaystyle \frac{p}{(2^\ell)^2} \;\;\sum_{\mathclap{{P \in \{I, Z\}^{\otimes \ell}}}} P \rho P^\dagger & \mbox{otherwise} .
    \end{cases}
\end{equation}
Finally, the twirled channel contains all combinations of $Z$ operators on each qubit
\begin{equation}
    \tilde{\mathcal{E}}_{C^{\ell-1} Z}^{\mathrm{decay}}(\rho) = \left(1 - \frac{(2^\ell - 1)^2}{(2^\ell)^2} p\right) \rho + \frac{2^\ell - 1}{(2^\ell)^2} p \sum_{\mathclap{\substack{P \in \{I, Z\}^{\otimes \ell} \\ P \neq I^{\otimes \ell}}}} P \rho P^\dagger .
\end{equation}
This result generalizes the one in Ref.~\cite{sahay2023} and is equivalent to \cref{eq:channels-CnZ} with $p_\ell = (2^\ell - 1)^2 p/ (2^\ell)^2$.

This would suggest $p_{3}\approx 1.36 \; p_{2}$. However, the gates have different decay probabilities. Then, accounting for the fact that a $CCZ$ gate is approximately twice as long as a $CZ$ gate~\cite{jandura2022a}, and taking into account extra contributions from laser imperfections, we choose $p_3 = 4 p_2$ for our numerical simulations, matching recent experiments~\cite{evered2023, bluvstein2023}.

\subsection{Idling errors}
\label{app:noise-model:idling}
In neutral-atom qubits, dephasing times $T_2$ ranging from a few to hundreds of milliseconds have been observed.
These can be extended to $T_2 \gtrsim \SI{1}{\second}$ with dynamical decoupling~\cite{bluvstein2022, graham2023, huie2023, norcia2023}, with recent demonstrations achieving $T_2 > \SI{10}{\second}$~\cite{barnes2022, manetsch2024}.
Conversely, the relaxation time $T_1$ spans between a few and hundreds of seconds ~\cite{bluvstein2022, ma2023, huie2023}.

As both quantum gates and shuttling over a few lattice spaces take significantly less time, a few microseconds at most, we neglect idling errors occurring during gates.
On the other hand, midcircuit measurements require times of the order of $\SIrange{1}{20}{\milli\second}$~\cite{bluvstein2023, norcia2023, graham2023, lis2023, huie2023, singh2023}. 
This leads to idling errors during measurements which cannot be neglected. 
For the following, we assume dephasing dominates, as $T_2 \ll T_1$.
Therefore, during measurements, every data qubit is subject to the channel
\begin{equation}
    \mathcal{E}_{\mathrm{idle}} (\rho) = \left(1 - p_{\mathrm{idle}} \right) \rho + p_{\mathrm{idle}} Z \rho Z,
\end{equation}
with
\begin{equation}
    p_{\mathrm{idle}} = \frac{1}{2} \left(1 - e^{-t/T_2}\right),
\end{equation}
where $t$ represents the time taken by measurements.

With a feed-forward time of $t \approx \SI{0.9}{\milli\second}$ and coherence $T_2 > \SI{1}{\second}$ from state-of-the-art QEC experiments with rubidium atoms~\cite{bluvstein2023}, $p_{\mathrm{idle}} \lesssim 10^{-3}$ was estimated.
Very recent results with arrays of cesium atoms observe a $T_2$ time that is an order of magnitude larger~\cite{manetsch2024}.
With further experimental improvements, an optimistic scenario for the future would be $t = \SI{0.1}{\milli\second}$ and $T_2 = \SI{50}{\second}$, leading to negligible idling rates ($p_{\mathrm{idle}} \approx 10^{-6}$).

\section{Additional numerical results}
\label{app:additional-numerics}

In this appendix, we report on the numerical simulations of our protocols under the simplistic noise model derived in \cref{sec:noise-model} for neutral-atom arrays, in the absence of idling errors.
This is shown in~\cref{fig:plogs_RDN}.

Results are qualitatively similar to the depolarizing noise (cf.~\cref{sec:depolarising-noise}), but the pseudo-thresholds are higher.
In particular, the pseudo-threshold of multiple implementations is found to be within the reach of current state-of-the-art experiments~\cite{bluvstein2023,finkelstein2024}, from which one can estimate a $p_2$ in the range of $\SIrange{0.005}{0.007}{}$.

\bibliography{library}

\end{document}